\documentclass[aps,pre,twocolumn,superscriptaddress,10pt]{revtex4-2}
\usepackage{graphicx} 
\usepackage{booktabs}
\usepackage{braket}
\usepackage{amsmath}
\usepackage{mathtools}
\usepackage{amssymb}
\usepackage{xcolor}
\usepackage[T1]{fontenc} 
\usepackage[utf8]{inputenc}
\newcommand{\bi}[1]{Fig.~\ref{f:#1}}
\newcommand{\lr}[1]{\langle #1 \rangle}
\newcommand{\e}[1]{Eq.~(\ref{eq:#1})}
\newcommand{\be}{\begin{equation}} \newcommand{\ee}{\end{equation}}
\newcommand{\ba}{\begin{eqnarray}} \newcommand{\ea}{\end{eqnarray}}
\usepackage[colorlinks=true, linkcolor=blue, citecolor=blue, urlcolor=black]{hyperref}

\begin{document}

\title{Effects of inertia on the asynchronous state of a disordered Kuramoto model}
\author{Yagmur Kati}
\email{yagmur.kati@hu-berlin.de}
\affiliation{Department of Physics, Humboldt-Universität zu Berlin, Newtonstr 15, 12489 Berlin, Germany}
\affiliation{Bernstein Center for Computational Neuroscience, Haus 2, Philippstr 13, 10115 Berlin, Germany}

\author{Ralf Toenjes}
\affiliation{Department of Physics, Humboldt-Universität zu Berlin, Newtonstr 15, 12489 Berlin, Germany}
\author{Benjamin Lindner}
\affiliation{Department of Physics, Humboldt-Universität zu Berlin, Newtonstr 15, 12489 Berlin, Germany}
\affiliation{Bernstein Center for Computational Neuroscience, Haus 2, Philippstr 13, 10115 Berlin, Germany}

\date{\today}

\begin{abstract}
We investigate the role of inertia in the asynchronous state of a disordered Kuramoto model. We extend an iterative simulation scheme to the case of the Kuramoto model with inertia in order to determine the self-consistent fluctuation statistics, specifically, the power spectra of network noise and single oscillators. Comparison with network simulations demonstrates that this works well whenever the system is in an asynchronous state. We also find an unexpected effect when varying the degree of inertia: the correlation time of the oscillators becomes minimal at an intermediate mass of the oscillators; correspondingly, the power spectra appear flatter and thus more similar to white noise around the same value of mass. We also find a similar effect for the Lyapunov spectra of the oscillators when the mass is varied.
\end{abstract}
\maketitle

\section{Introduction}
The Kuramoto model has been extensively studied for its ability to describe synchronization in coupled oscillators \cite{Kur84,PikRos01} (see, specifically, the review \cite{AceBon05} and the many references therein). However, less attention has been paid to the asynchronous states, where oscillators exhibit incoherent dynamics \cite{StiRad98,KatRan24,PruEng24}. Understanding these asynchronous regimes is essential for many real-world systems characterized by disordered interactions and diverse natural frequencies, such as neural networks \cite{VanSom96,RenDel10,Ost14}, coupled hair cells in the cochlea \cite{TalTub98,KuEll09,FruJue14,WitBel24},  and power grids \cite{WitHel22} to name but a few examples from different areas. In particular, the inclusion of inertia in the model significantly alters the oscillator dynamics, but its effects in the asynchronous regime remain underexplored.

Inertia introduces complexity to the system by modifying the oscillator response to coupling and disorder, affecting both their time evolution and spectral properties. While synchronization in inertial systems has been previously studied, little focus has been placed on how inertia influences the detailed statistical structure of the asynchronous state, in which oscillators are uncorrelated among each other but still possess a nontrivial temporal correlation. The latter correlations are crucial for understanding how complex networks function under weak coupling and strong disorder conditions.

A great advantage of the asynchronous state is that here stochastic mean-field theories can be used to determine the self-consistent fluctuation statistics, in particular the aforementioned temporal correlations characterized by correlation functions or, equivalently, power spectra. Because every oscillator in the network is a driven element but also a driver of others, the statistics of input fluctuations to each oscillator are connected by a simple self-consistency condition to the output statistics of each oscillator. This approach has been pursued for networks of rate units in the work of Sompolinsky et al. \cite{SomCri88} (generalized to heterogeneous network settings in \cite{AljSte15,KadSom15,MasOst17}), for networks of sparsely connected spiking neurons \cite{LerSte06,DumWie14,PenVel18,VelLin19} and for networks of coupled rotators \cite{VanLin18,RanLin22,RanLin23}. For the  Kuramoto network with random connectivity and without inertia the method was first developed by Stiller and Radons in the 1990's \cite{StiRad98} and applied for the determination of self-consistent power spectra in the asynchronous state by two of us and a collaborator \cite{KatRan24}. It has not been applied so far to a Kuramoto network with inertia to the best of our knowledge. 

In this work, we investigate the asynchronous states of the Kuramoto model with inertia, extending the iterative mean-field (IMF) method to compute the power spectra of individual oscillators.
By comparison to numerical simulations of large networks, we demonstrate that the IMF method works for the full range of inertia deep into the underdamped regime. We furthermore uncover a nontrivial  
effect of inertia in this model: Network fluctuations are temporally least correlated for an intermediate value of the oscillator mass and the power spectra widen around zero, i.e. the network fluctuations cover a broader frequency range, more similar to white noise. This maximum in temporal disorder occurs when a transition from monotonic decay in the autocorrelation function to a damped  oscillation is observed. Correlation times of fluctuations are relevant for signal processing by neural networks (see e.g. \cite{Ost14}) and short correlation times might be especially favorable or disadvantageous, depending on the specific task at hand.

Our paper is organized as follows. In the next section, we introduce the model and the spectral statistics of interest. In sec.~\ref{sec:IMF}, we sketch the iterative mean-field (IMF) method that can be used to determine  the self-consistent correlation statistics from repeated simulations of a single oscillator. In sec.~\ref{sec:results}, our results section, we  first demonstrate that the IMF method works for the Kuramoto model with inertia. We then show a remarkable effect when varying the oscillator mass: when the frequency disorder is absent or small, the spectral width becomes maximal and the correlation time becomes minimal at an intermediate mass. The self-consistent temporal network correlations are least pronounced for a mass at which the correlation function changes from a monotonic shape to that of a damped oscillation. Finally, we inspect the Lyapunov spectrum and the Kolmogorov-Sinai entropy for the network for varying oscillator mass and find that at the intermediate mass that minimizes the correlations, the Kolmogorov-Sinai entropy for the network is maximal. We conclude in  sec.~\ref{sec:conclusions} with a brief summary of our results and an outlook to open problems.
\section{Model and measures}
\subsection{Kuramoto model with inertia and random coupling parameters}
We investigate the temporal statistics in a system of randomly coupled Kuramoto phase oscillators with inertia, where the phase  $\theta_\ell$ of each oscillator evolves as 
\begin{equation}\label{eq:eom}
m \ddot{\theta}_\ell  = \gamma(\omega_\ell-\dot{\theta}_\ell) + \sum_{m=1}^N K_{\ell m} \sin(\theta_m - \theta_\ell).
\end{equation}
Here $m$ and $\gamma$ represent a mass and a friction coefficient, chosen identical for all oscillators, $\omega_\ell$ are the natural frequencies, and the matrix $K_{\ell m}$ quantifies the coupling between oscillators. Phases are attracted for positive coupling coefficients $K_{\ell m}$, while negative values represent repulsive coupling.
The natural frequencies $\omega_\ell$ are Gaussian distributed as
\begin{equation}
\omega_\ell = \mathcal{G}_\ell \sigma_\omega, \quad \langle \omega_\ell \rangle = 0, \quad \langle \omega_\ell \omega_m \rangle = \sigma_\omega^2 \delta_{\ell m},
\end{equation}
where the $\mathcal{G}_\ell$ are independent Gaussian random numbers with zero mean and unit variance. The equations of motion are invariant under a transformation into a co-rotating reference frame $\theta\to\theta+\Omega t$, $\dot\theta\to\dot\theta+\Omega$ and $\omega_\ell \to\omega_\ell+\Omega$ so that the mean natural frequency may be set to zero. Similarly, the coupling coefficients $K_{\ell m}$ are taken as independent Gaussian random variables,
\begin{equation}
    K_{\ell m} = \frac{K}{N} + \frac{k \mathcal{G}_{\ell m}}{\sqrt{N}}
\end{equation}
where $\mathcal{G}_{\ell m}$ represents a Gaussian matrix with independent entries, each possessing zero mean and unit variance. 
The parameter $K$ is the effective coupling strength to the mean field, while $k$ quantifies the coupling heterogeneity. Using independent random heterogeneities simplifies the dynamics and is justified as an approximation in sparse, directed coupling networks.
After dividing the equations of motion by $\gamma$ and rescaling the time in units of $\tau=k/\gamma$ we can set the friction $\gamma$ as well as the coupling heterogeneity $k$ to one, provided both are nonzero. We use $\gamma=1$ in the following unless explicitly noted otherwise. 

In contrast to other studies, we are not interested in the transition to synchronization but in the temporal statistics of the dynamic fluctuations in the incoherent state, where the phases are distributed uniformly. Even without frequency disorder, i.e. $\sigma_\omega=0$, the mixture of random attractive and repulsive coupling prevents synchronization. In the incoherent state fluctuations of the total inputs to the single oscillators (below defined by the local mean fields $\zeta_\ell(t)$) are statistically independent and of the order $k$, independent of the system size, while the global mean field 
\begin{equation}
    re^{i\Theta} = \frac{1}{N}\sum_{\ell=1}^N e^{i\theta_\ell}
\end{equation}
is negligible of the order $K/\sqrt{N}$. Upon increasing $K$ the system will go through a phase transition in the limit $N\to\infty$. The incoherent state becomes unstable and global collective oscillations of a nonzero mean field emerge \cite{AceBon05}. The order of synchronization is measured by the Kuramoto order parameter $r$, the modulus of the complex global mean field which takes values between $r=0$ ($O(1/\sqrt{N}$) in the incoherent state, $0<r<1$ in states of partial synchronization, and $r=1$ for complete synchronization, i.e. $\theta_\ell=\Theta$ for identical oscillators. 

We integrate the equations of motion \eqref{eq:eom} from uniformly random distributed  initial phases and zero phase velocities using the Runge-Kutta method. For all results, we discard a transient time $t_d$ after which the variance of the phase velocities becomes stationary, ensuring that the system reaches a steady state. To test whether we are in the asynchronous state, we then compute the time-averaged order parameter $r$ over a time window $T$. In large systems, the averaged statistics should become independent of the disorder realization in $\omega_\ell$ and $K_{\ell m}$. We will explicitly note where we take additional averages over different disorder realizations.
\begin{figure}[h!]%
\includegraphics[width=0.45\textwidth]{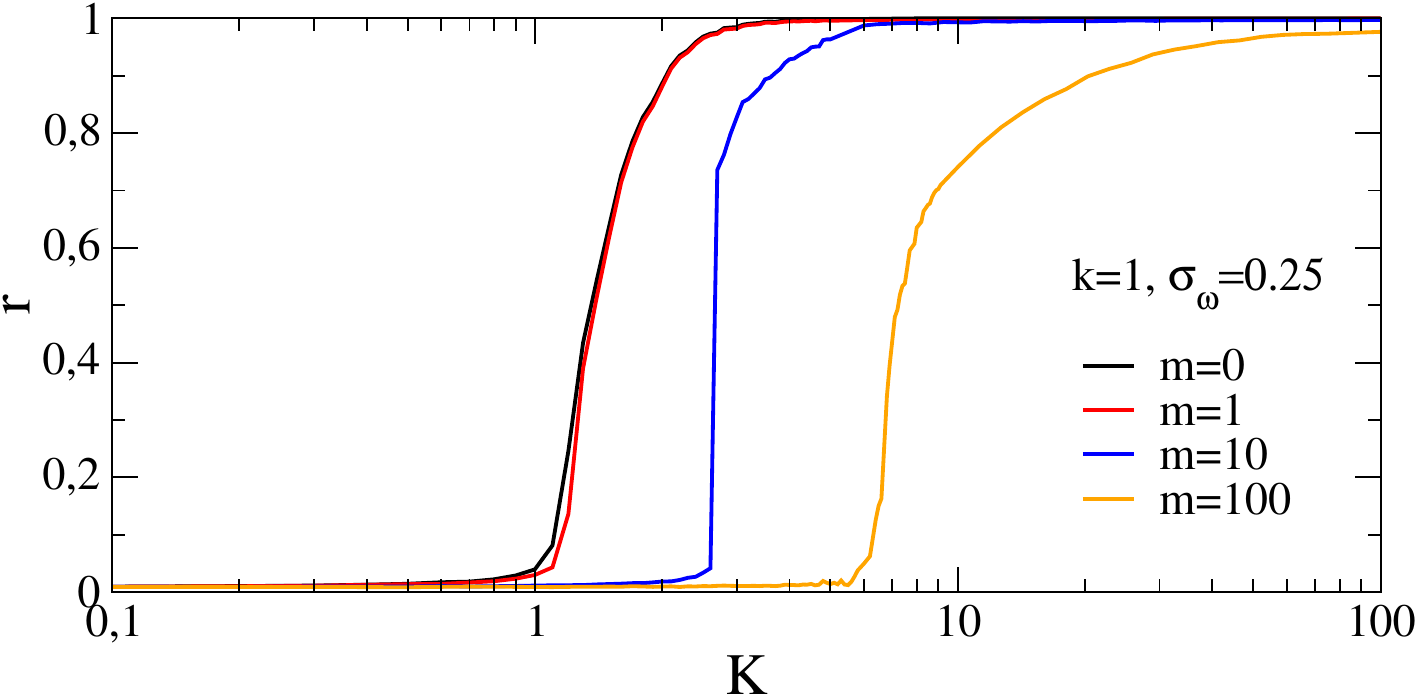} 
\caption{\textbf{Defining the asynchronous regime.} The order parameter $r$ vs $K$ for a system with a random network ($k=1$) or a random frequency ($\sigma_\omega=0.25$) distribution for $m=0, 1, 10, 100$. Other parameters: $N=10^4, T=2000, t_d=200, R=1$. For all masses shown and $K<1$, the network is in the asynchronous regime with $r\ll 1$.
}
\label{f:orderparam}
\end{figure}

In \bi{orderparam} we show the order parameter for a situation with both frequency and coupling disorder ($\sigma_\omega=0.25,k=1$) as a function of the mean coupling strength $K$ for different values of mass. 
Increasing inertia has two effects: i) it leads to an enlargement of the range of mean coupling coefficient with an asynchronous solution ($r\ll1$), i.e. it pushes the critical coupling $K_c$ to higher values; ii) in the strongly underdamped case, the second-order phase transition turns into a first-order phase transition (note the jump in the curve for $m=10$), which is in line with what has been reported before \cite{TanLic97,Sabhahit24}.  

In any case, for all inspected values of the mass, the asynchronous state ($r\ll 1$) always exists over a finite range of mean coupling strengths. In this parameter region the temporal statistics of the dynamics is independent of $K$. In the remainder of the study, we will set $K=0$.

\subsection{Power Spectrum Calculation}

To further analyze the fluctuations in the incoherent regime, we compute the power spectrum of individual oscillators, denoted as $S_{x_\ell}(\omega)$, where $x_\ell = e^{i\theta_\ell}$. The power spectrum is defined as:
\begin{equation}
S_{x_\ell}(\omega) = \lim_{T \to \infty} \frac{1}{T} \left\langle |\tilde{x}_\ell(\omega)|^2 \right\rangle, \quad \tilde{x}_\ell(\omega) = \int_0^T  e^{i\omega t} x_\ell(t)\,dt,
\end{equation}
where $\tilde{x}_\ell(\omega)$ is the Fourier transform of $x_\ell(t)$ computed over a finite time interval $T$. Common temporal statistics are characterized by the averaged power spectrum,
\begin{equation}
S_z(\omega) = \frac{1}{N} \sum_{\ell=1}^N S_{x_\ell}(\omega).
\end{equation}
We obtain these spectra by integrating the full set of second-order differential equations (\e{eom}) using a fourth-order Runge-Kutta method with a relative tolerance of $10^{-8}$. We refer to this direct numerical integration as the network dynamics (ND) method and use it as a benchmark against the iterative mean-field (IMF) method introduced in the following section. In \bi{Fig2ab} and \bi{Fig3abcd}, the ND results are shown as black dashed lines for comparison with the IMF method. 

The simulation is run for a total time period $T$ (with $T=10^5$ for \bi{Fig2ab} and $T=10^4$ otherwise), discarding the initial time $t_d=1000$ to eliminate transients. To allow oscillator-resolved analysis, the natural frequencies $\omega_\ell$ are drawn from the prescribed distribution and held fixed, as is the Gaussian-distributed connectivity matrix $K_{\ell m}$. The system size is chosen as $N=10^4$ to minimize dependence on the specific realization of the connectivity matrix $K_{\ell m}$, which is also drawn from a Gaussian distribution.

The power spectrum is computed using the fast Fourier transform (FFT). The single-oscillator power spectrum $S_{x_\ell}(\omega)$ is evaluated for a selected oscillator ($\ell=3$ in our case), while the mean power spectrum $S_z(\omega)$ is obtained by averaging over all oscillators:
\begin{equation}
S_z(\omega) = \frac{1}{N} \sum_{\ell=1}^{N} |\tilde{x}_\ell(\omega)|^2.
\end{equation}

The results are smoothed by binning neighboring frequency points to improve spectral accuracy. We divided the frequency axis into $M=2000$ uniform bins, with $T/(M dt)$ frequency points averaged per bin using an accumulation method to reduce statistical noise.

\section{Iterative Mean Field Method}
\label{sec:IMF}
In this study, we employ the iterative mean field (IMF) method, often also referred to as dynamic mean field method \cite{CriSom18},  to analyze the asynchronous states of the Kuramoto model with inertia and disorder in coupling and natural frequencies. This method provides a self-consistent approach to compute the power spectra of single oscillators, approximating the network noise experienced by each oscillator as a Gaussian process, iteratively refined until convergence. The method has been used for spin systems in the early 1990's \cite{EisOpp92}, for networks of spiking neurons \cite{LerSte06,DumWie14,PenVel18}, and specifically for the Kuramoto model  with disorder in the coupling coefficients (but without inertia) in \cite{StiRad98,KatRan24,PruRos24,PruEng24}.  

For our system with inertia, we start from the equations of motion for the oscillators \e{eom} and express the coupling term in the dynamics as a multiplicative network noise:
\begin{align}
    m \ddot{\theta}_\ell(t) + \dot{\theta}_\ell(t)  = \omega_\ell + \text{Im}\left( e^{-i\theta_\ell(t)} \zeta_\ell(t) \right),
\end{align}
where the network noise $\zeta_\ell(t)$ is defined as:
\begin{equation}\label{eq:NWnoise}
    \zeta_\ell(t) = \sum_{m=1}^N K_{\ell m} e^{i\theta_m(t)}.
\end{equation}
For large networks in the asynchronous regime (excluding the Volcano transition case observed with symmetric coupling coefficients, see \cite{Dai93,PruRos24}), this noise behaves as a superposition of independent Gaussian processes due to the central limit theorem. The autocorrelation of the network noise can be written as:
\begin{align} 
\left\langle \zeta_\ell(t) \zeta_\ell (t') \right\rangle &=
\left \langle K_{\ell m} K_{\ell n} e^{i(\theta_m(t')-\theta_n(t))} \right\rangle \nonumber \\
&= \sum_{m,n} \left(\frac{k^2}{N} \delta_{mn} + \frac{K^2}{N^2} \right) \left\langle e^{i(\theta_m(t') - \theta_n(t))} \right\rangle. 
\end{align} 
Assuming uncorrelated oscillators yields
\begin{align} \left\langle e^{i(\theta_m(t') - \theta_n(t))} \right\rangle = \delta_{mn} \left\langle e^{i(\theta_m(t') - \theta_m(t))} \right\rangle, \end{align} 
leading to the simplified form
\begin{align} \label{eq:zeta1}
\left\langle \zeta(t) \zeta (t') \right\rangle &= \left(k^2 + K^2/N\right) \left\langle e^{i(\theta(t')-\theta(t))} \right\rangle. 
\end{align}
The right-hand side thus represents the autocorrelation of a single oscillator’s phase pointer. 
In the Fourier domain, the network noise spectrum $S_\zeta(\omega)$ is proportional to the power spectrum of a single oscillator averaged over the network:
\begin{equation} \label{eq:spec_zeta}
S_\zeta(\omega) = (k^2 + K^2/N) S_z(\omega), 
\end{equation} 
which demonstrates the self-consistent relationship between individual oscillator fluctuations and the emergent network noise. 
In the large-$N$ limit, the system simplifies to a single effective oscillator described by: 
\begin{align}\label{eq:eomIMF}
    m \ddot{\theta}(t) + \dot{\theta}(t) 
   =  \sigma_\omega \xi_\omega + \text{Im}\left( e^{-i\theta(t)} \zeta(t) \right).
\end{align}
Here, the static Gaussian noise $\xi_\omega$ has zero mean and unit variance, $\langle \xi_\omega(t) \xi_\omega(t') \rangle = 1$, while the network noise $\zeta(t)$ is a complex-valued stochastic process with zero mean. Unlike $\xi_\omega$, which is predefined and directly sampled in \e{eomIMF}, $\zeta(t)$ must be determined self-consistently via \e{zeta1} or its Fourier counterpart \e{spec_zeta}, as it is shaped by the dynamics it drives.
  
The iterative mean-field (IMF) method starts with an initial guess for the network noise spectrum $S_\zeta(\omega)$, typically a Lorentzian. Using this initial spectrum, we generate surrogate noise samples in Fourier space as  
\begin{equation}\label{eq:tildezeta}
    \tilde{\zeta}(\omega) = (\mathcal{G}_r(\omega) + i \mathcal{G}_i(\omega)) 
    \sqrt{\frac{S_\zeta(\omega) T}{2}},   
\end{equation}  
where $\mathcal{G}_r(\omega)$ and $\mathcal{G}_i(\omega)$ are independent Gaussian random numbers with zero mean and unit variance. The inverse Fourier transform provides a realization of $\zeta(t)$, which is then used to simulate \eqref{eq:eomIMF} across multiple trials.  

For each frequency $\sigma_\omega \xi_\omega$, we generate $R$ independent trials with different noise samples, resulting in $M_{\text{trial}} = N R$ trials across $N$ frequencies. With a large system size $N=10^4$, we set $R=1$ since $M_{\text{trial}}$ is sufficient for averaging, and we fix the Gaussian-distributed natural frequencies. We used the exact frequency distribution as in the ND method to compare spectra for the same selected oscillator directly.  

From these simulations, where we solve \e{eomIMF} with \e{tildezeta}, we compute the power spectrum of the phase pointer $e^{i\theta(t)}$ and update $S_\zeta(\omega)$ via \e{spec_zeta}. The updated spectrum is then used to generate new noise samples, and the process repeats until convergence.  $S_\zeta(\omega)$ stabilizes within $I=20$ iterations for our parameters, after which the mean spectrum is given by $S_z(\omega) = S_\zeta(\omega) / (k^2 + K^2/N)$. 

To compute single-oscillator spectra $S_{x_\ell}(\omega)$, we use the converged $S_\zeta(\omega)$ in \e{tildezeta}, select a natural frequency $\omega_\ell = \sigma_\omega \xi_\omega$, and solve \e{eomIMF} with \e{tildezeta} for $10^4$ trials without further iterations. The power spectrum $S_{x_\ell}(\omega)$ is then obtained via Fourier transform of $e^{i\theta(t)}$. We use the Euler-Maruyama method with a time step $dt=0.05$ for time integration and fast Fourier transform for spectral analysis in all IMF results. The code is available in \cite{Kati2025Power} for further details.

\section{Results}
\label{sec:results}
\subsection{Validation of the IMF method for the Kuramoto system with inertia}
We test the IMF method for various values of the mass and compare the resulting power spectra to network simulations; results are shown in \bi{Fig2ab}. We consider a situation with mild frequency disorder, pick one of the oscillators, say the third one with a frequency of $\omega_3=0.0165$ and show its power spectrum in \bi{Fig2ab}a. Because for the single oscillator, we have to rely on a pure time average for computing the power spectrum, the network spectrum of the third oscillator is not very smooth but shaped by measurement noise (cf. dashed lines). While the results for time periods $T=10^4$ and $T=10^5$ were indistinguishable for IMF method, the single oscillator full network dynamics in \bi{Fig2ab}a required $T=10^5$ to have results smooth enough to be compared with IMF on linear scale. Note that the CPU time for ND is 10 times longer than IMF if we use $T=10^5, N=10^4, R=1$ for both methods. For the IMF method, we can easily simulate many realizations and obtain much smoother estimates of the spectra (cf. colored solid lines) that agree well with the network simulations. The spectra are centered around the oscillator's eigenfrequency  $\omega_3=0.0165$ but change their shape considerably when the oscillator mass is changed.
\begin{figure}[h!] 
\centering
\includegraphics[width=0.4\textwidth]{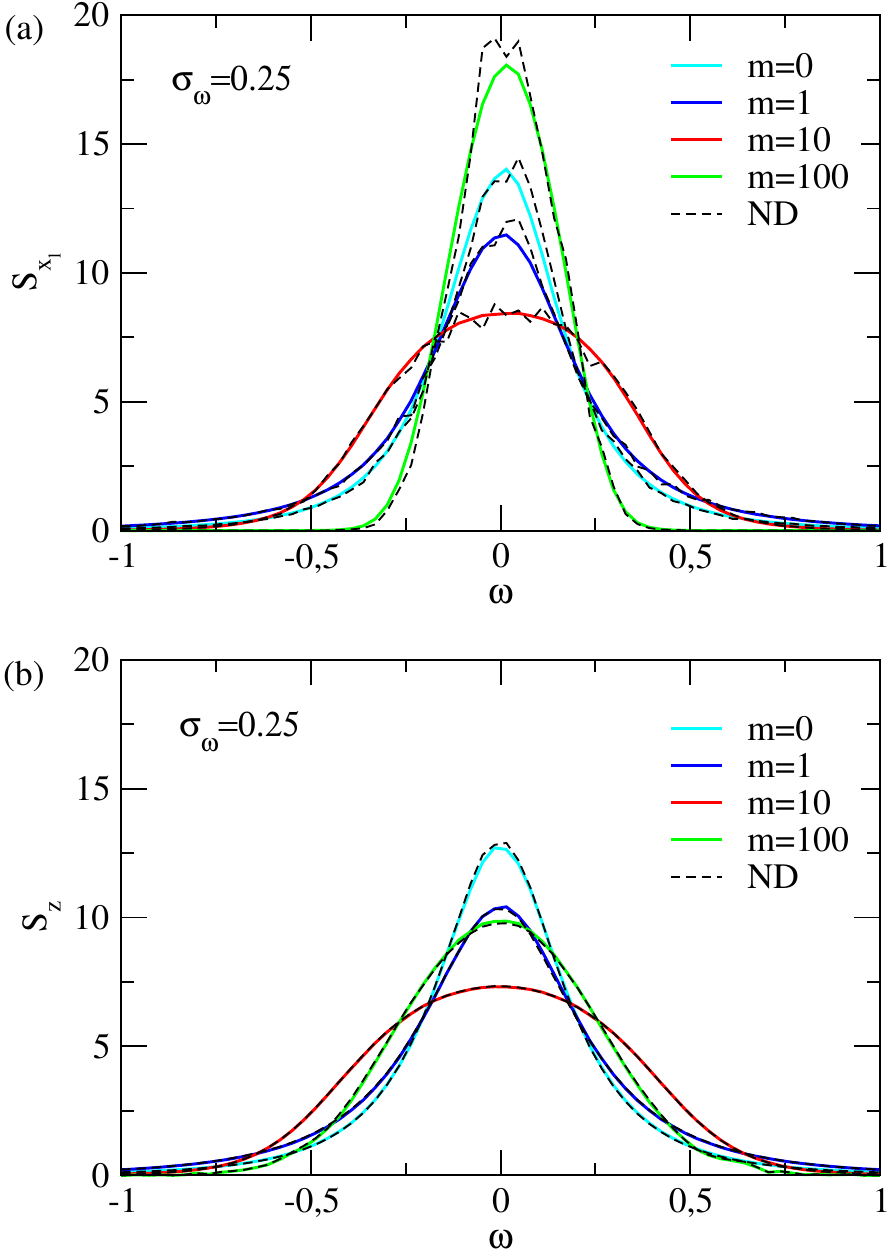}
\caption{\textbf{IMF method works for the Kuramoto model with inertia.}  
A comparison between the IMF and ND methods is presented for masses $m=0, 1, 10, 100$. (a) Single-oscillator spectra for the selected oscillator with $\omega_3=0.0165$ and (b) mean power spectra are shown. The IMF results are represented by solid lines, while the ND results are shown with black dashed lines. Fixed parameters: $\sigma_\omega=0.25, k=1, N=10^4, T=10^5, t_d=1000, R=1$.}  
\label{f:Fig2ab} 
\end{figure}

The similarity of the results of the network and IMF methods can be even more appreciated for the much smoother network-averaged power spectra $S_z(\omega)$ displayed in \bi{Fig2ab}b. The average spectrum is centered and symmetric around zero because of the symmetric (Gaussian) distribution of the natural frequencies $\omega_\ell$. The agreement between the network simulation results and the solution of the self-consistent model is excellent. In particular, no discrepancy emerges with an inclusion of inertia effects; the IMF method works even deep in the underdamped regime. 

We have found that the self-consistent scheme yields power spectra equivalent to those resulting from network simulations (not shown) for stronger frequency heterogeneity ($\sigma_\omega=2$) or in the absence of heterogeneity ($\sigma_\omega=0$), for smaller or larger strength of coupling disorder  (not shown). This agreement also holds for different values of the mean coupling constant $K$ as long as $|K| < K_c$ and for system sizes as small as $N = 8$ (as shown in our previous work \cite{KatRan24}).

\subsection{Minimal correlation time at intermediate mass}
We now inspect the shape change of the spectra with increasing inertia in more detail, looking again at the spectra in \bi{Fig2ab}.  

Remarkably, coming from $m=0$ (cyan line in panel a), the first effect of a small but finite inertia is a broadening of the spectral peak for $m=1$ (blue line). A further increase in mass to $m=10$ (red line) yields a rather broad plateau for frequencies between -0.3 and 0.3; here the oscillator spectrum is almost like a band-pass limited white noise in the sense that all frequencies in the mentioned broad band contribute equally to the system's variance. A further increase, however, to $m=100$ (green line) causes again a more narrow spectral peak, both observable in the single oscillator's (panel a) and the average spectra (panel b).  Apparently, there  seems to be an optimal value of mass with respect to the flatness (``whiteness'') of the oscillator spectrum.  
We can estimate the flatness by the spectral width $\Delta\omega$, for which we use here the differences of frequencies at which the spectrum has decayed to 80\% of its maximum value $S_z(\omega_{\text{max}})$, i.e.
\be
\Delta \omega=\omega_R-\omega_L, \;\; \text{with}\;\; S_z(\omega_R)=S_z(\omega_L)=0.8\,S_z(\omega_\text{max}).
\label{eq:delta_omega}
\ee
If the oscillator generates a signal that is close to white noise, we would expect its correlation function to have a minimal time scale. Put differently, the maximal flatness of the spectrum may correspond to a minimum of the system's correlation time, which can be defined in different ways, for instance, as 
\be
\hat{\tau}_C=\int_0^\infty d\tau C(\tau)/C(0), 
\label{eq:tauc}
\ee
a definition that is particularly appropriate for a monotonically decaying and positive correlation function. Numerically, we compute $C(\tau)$ using an inverse fast Fourier transform of the power spectrum, following the Wiener–Khinchin theorem:
\be
C(\tau) = \int_{-\infty}^{\infty}   S_z(\omega)e^{-i\omega\tau}\,d\omega. 
\ee
Since $C(\tau)$ is symmetric, we extract the positive-delay part, $\tau \geq 0$, for analysis, as shown in \bi{Fig3abcd}b.

\begin{figure}[h!]
\centering
\includegraphics[width=0.4\textwidth]{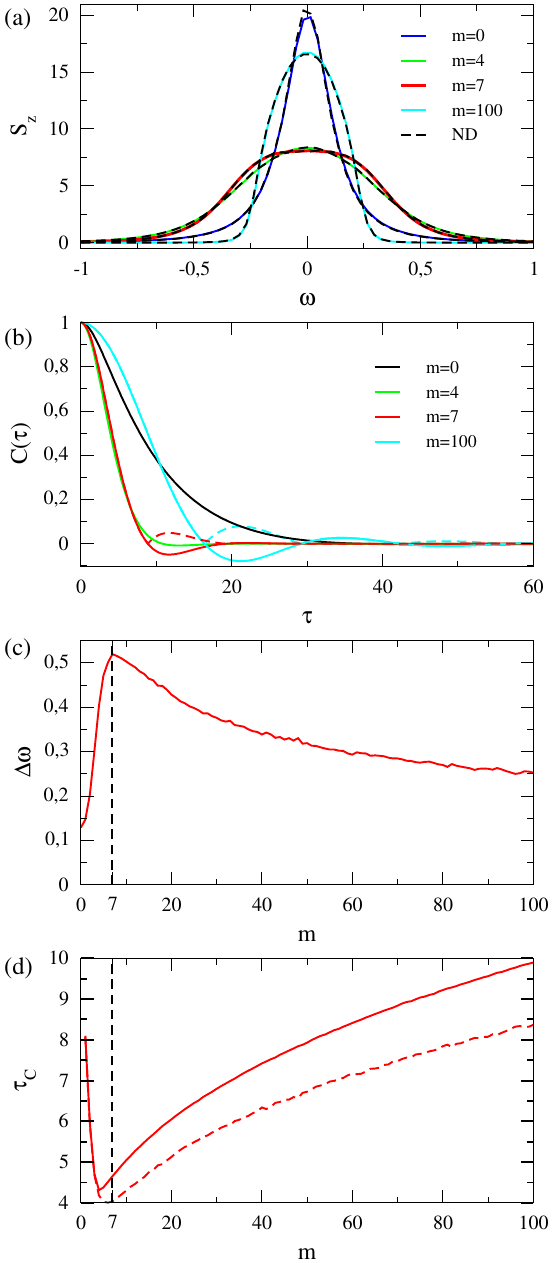}
\caption{\textbf{Maximized spectral flatness and minimized temporal correlation for an intermediate oscillator mass.} 
(a) Mean power spectrum for  $m=0, 4, 7, 100$: solid lines (IMF method), black dashed lines (ND method). (b) Autocorrelation function  for $m=0, 4, 7, 100$. Dashed lines, matching the colors of the corresponding solid lines, represent $|C(\tau)|$ and highlight its use in (d). (c) Spectrum width, \e{delta_omega} at 80\% maximum of $S_z(\omega)$ vs mass with a maximum  at $m=7$, marked by a black dashed line. (d) Correlation time vs mass: $\tau_C$ according to \e{tauc} has a minimum at $m= 4$. Dashed red line represents $\hat{\tau}_C=S_z(0)/2$, according to \e{tauc_absval}, with a minimum at $m=7$, marked by a black dashed line. Fixed parameters: $k=1, \sigma_\omega=0, N=10^4, I=20, R=1, T=10^4, t_d=1000$.}
\label{f:Fig3abcd}
\end{figure}

\begin{figure}[h!]
\centering 
\includegraphics[width=0.4\textwidth]{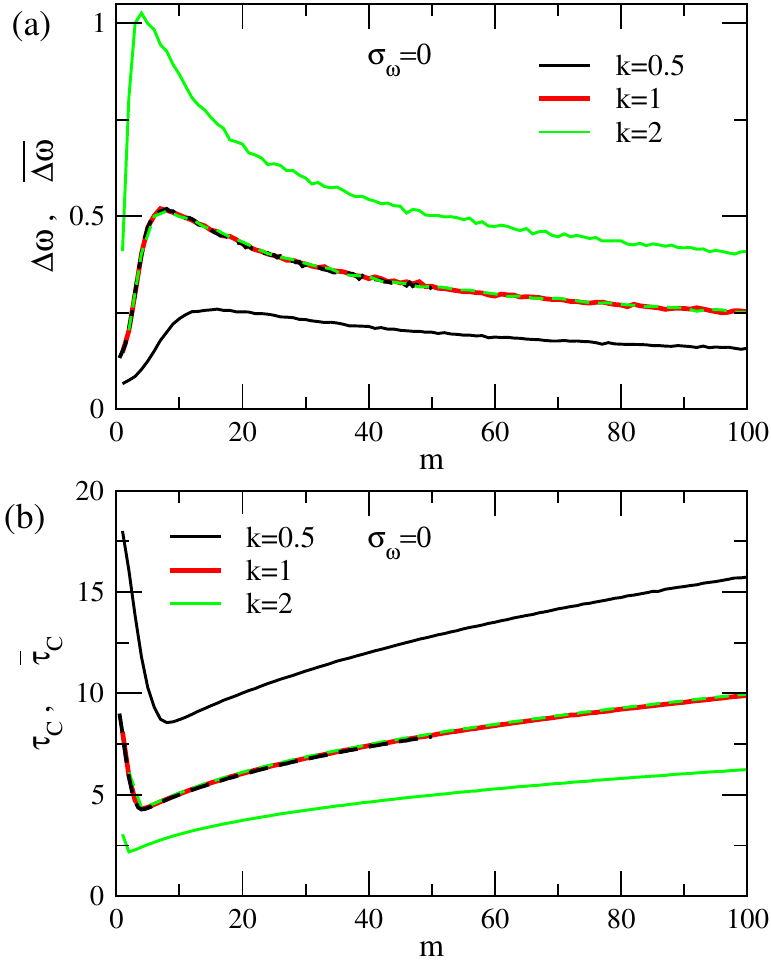}
\caption{\textbf{Dependence on the coupling coefficient can be captured by scaling arguments.} 
(a) Spectrum width at 80\% maximum of $S_z(\omega)$ for $k=0.5, 1, 2$. Solid lines represent $\Delta\omega$, and dashed lines represent the rescaled $\overline{\Delta\omega}$ using \e{rescale_k}. 
(b) Correlation time as a function of mass for  $k=0.5,1,2$. Dashed lines show the rescaled $\tau_c$ according to \e{rescale_k}.  
Fixed parameters: $\sigma_\omega=0, N=10^4, R=1, T=10^4, t_d=1000$. }
\label{f:Fig4ab}
\end{figure}

Applying the correlation-time definition (\ref{eq:tauc}) to the pointers $x_\ell=e^{i \theta_\ell(t)}$, we note that i) the variance of the function is always one, $\lr{x_\ell(t)x_\ell^*(t)}=C(0)=1$; ii) the integral can be regarded as half of a Fourier-transform at $\omega=0$. Consequently, the correlation time of the $\ell$th oscillator can be expressed by $\hat{\tau}_{C,\ell}=S_{x_\ell}(0)/2$. In Figs.~\ref{f:Fig3abcd}(d) and \ref{f:Fig4ab}(b), we show the mean correlation time for all oscillators  $\hat{\tau}_C= S_z(0)/2$. 

If we deal with correlation functions that show a (damped) oscillation, an alterative definition is 
\be
{\tau}_C=\int_0^\infty d\tau |C(\tau)/C(0)|. 
\label{eq:tauc_absval}
\ee
Note that, trivially, ${\tau}_C=\hat{\tau}_C$ for entirely positive correlation functions and, generally, ${\tau}_C\ge\hat{\tau}_C$. When plotting  ${\tau}_c$ and $\hat{\tau}_c$ vs a parameter, a separation of the curves indicates the emergence of an oscillation  (or at least an undershooting) in the correlation function. In \bi{Fig3abcd}d, we see that ${\tau}_c$ (solid line) and $\hat{\tau}_c$ (dashed line) are indistinguishable for $m\leq 4$, as the correlation functions for mass $m=0$ and $m=4$ in \bi{Fig3abcd}b are mainly positive and monotonic. The deviation between ${\tau}_c$ and $\hat{\tau}_c$ increases with growing mass due to oscillatory behavior in $C(\tau)$. Before discussing Figs.~\ref{f:Fig3abcd} and \ref{f:Fig4ab} in detail, we note that the data used to compute $\Delta\omega, \tau_C$, and $\tilde{\tau_C}$ were smoothed using a mild Savitzky-Golay filter, applied solely at this step. Although this minimal filtering did not affect the general trends observed, we emphasize that no filtering was used for the power spectra presented in this paper.

In \bi{Fig3abcd} we inspect spectra, correlation functions and the flatness and correlation characteristics $\Delta \omega$, $\tau_C$ and $\hat{\tau}_C$ for the case $\sigma_\omega=0$, $\gamma=1$, and $k=1$. The plot demonstrates that the minimization of correlations for a finite and non-vanishing mass exists if there is no variability in the natural frequencies of the oscillators. One numerical advantage of the case $\sigma_\omega=0$ is that we can average over the oscillators and obtain the same spectrum as for the single oscillator - only much smoother because of the averaging. 

Specifically, in \bi{Fig3abcd}a we display spectra for different masses that illustrate essentially the same effect of a maximized  spectral flatness, here at $m=7$. In panel (b), the corresponding correlation functions are shown (and their absolute value by dashed lines, i.e. the integrand of the correlation time $\tau_C$ in \e{tauc_absval}). Clearly, the correlation extends over the smallest time period for $m=7$. Consequently, the width of the spectrum (panel c) and the correlation time (panel d) display a maximum and minimum, respectively. The exact position of the minimum depends on the definition of the correlation time: integrating over the absolute value, the minimum is attained at $m\approx 4$, whereas the integral over the correlation function itself (i.e. half of the power spectrum) is minimized around $m=7$. Interestingly, both minima are close to the value of mass at which the curves for $\tau_C$ and $\hat{\tau}_C$ depart from each other, i.e. for the mass at which the correlation function starts showing a damped oscillation (or undershoots the time-lag axis).    

One may wonder how the effect depends on the network heterogeneity $k$. For the case of vanishing frequency heterogeneity ($\sigma_\omega=0$), we can rescale the dynamics in a simple manner  (cf. appendix \ref{app:rescaling}, \e{tauc_rescaled_gamma}, \e{del_om_rescaled_gamma} with $\gamma=1$). Stating the explicit dependence on the parameters $m$ and $k$, we obtain
\begin{equation}\label{eq:rescale_k}
\Delta \omega(k,m)=\frac{1}{k}\Delta{\omega}(mk,1),\;\; \tau_c(k,m)=k{\tau}_C(mk,1),    
\end{equation}
relations that are confirmed in \bi{Fig4ab} by plotting data for three different $k$ on top of one another (rescaling the $m$ and $\Delta \omega$ or $\tau_c$ axes according to the relations). 
The rescaling argument tells us, that a stronger coupling heterogeneity causes the maximum of the spectral width to move to larger masses and to become larger. Correspondingly, the minimum in the correlation time moves to smaller masses and becomes lower. For finite frequency disorder, the standard deviation of frequencies has to be rescaled as well when $k$ is changed (cf. appendix \ref{app:rescaling}, \e{sigma_om_rescaled_gamma}).

\begin{figure}[h!]
\centering 
\includegraphics[width=0.4\textwidth]{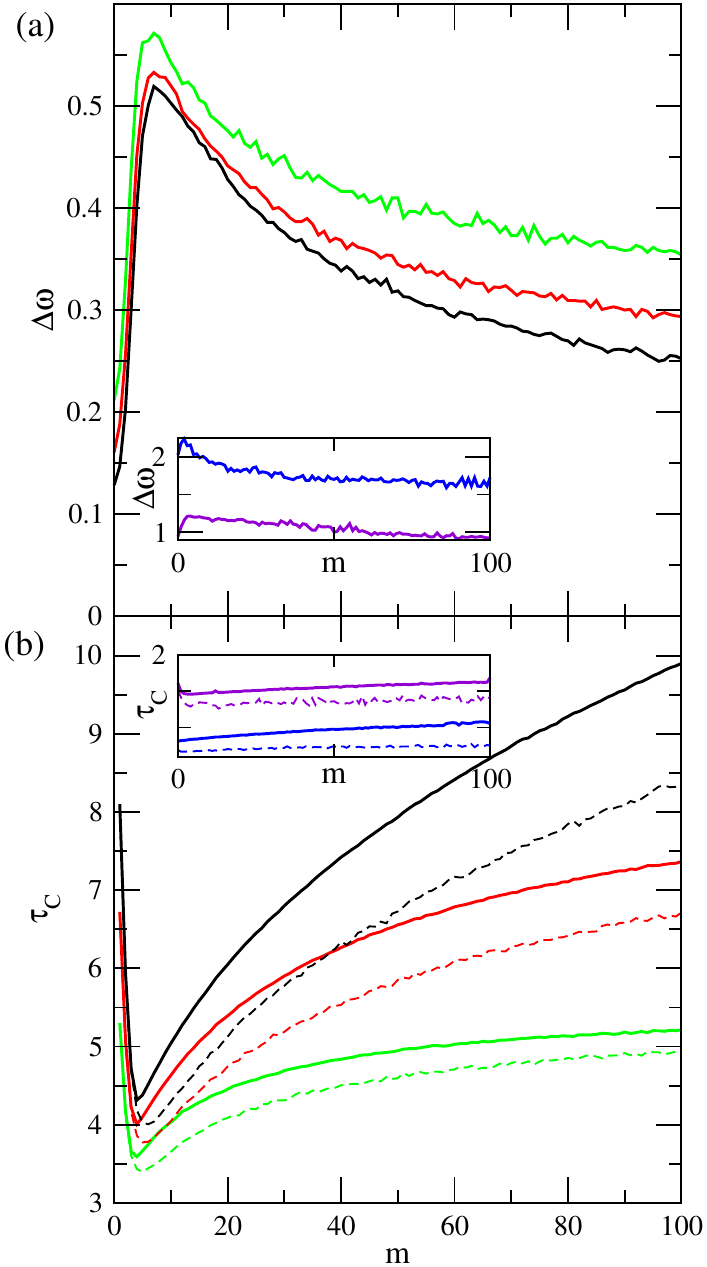}
\caption{\textbf{Maximizing flatness and minimizing correlation time at intermediate mass is robust against moderate 
disorder in natural frequencies}.      For $\sigma_{\omega} = 0, 0.15, 0.25$, the lines are black, red, and green, respectively, while the insets show $\sigma_{\omega} = 1$ (purple) and 2 (blue).  (a) Spectral width $\Delta\omega$ at 80\% maximum of $S_z(\omega)$ vs mass $m$. (b) Correlation time $\tau_{C}$ vs mass $m$. Dashed lines represent $S_z(0)/2$, with colors matching the solid lines for each corresponding $\sigma_\omega$. Fixed parameters: $k=1, N=10^4, R=1, T=10^4, t_d=1000$. }
\label{f:Fig5ab}
\end{figure}

Finally, we inspect the effect of an additional disorder in the natural frequencies on the observed whitening effect at intermediate mass. In \bi{Fig5ab} we see that the effect persists for moderate non-vanishing values of $\sigma_\omega$ and that an increase of frequency disorder leads to a further increase in the band width and decrease in correlation time. It turns out, however, that for very strong frequency disorder (larger than the disorder in coupling coefficients), the correlation time becomes so short already for $m=0$ that an increase in inertia cannot decrease the correlation time or increase the spectral width substantially (see blue lines for $\sigma_\omega=2$ in \bi{Fig5ab} insets).

\subsection{Lyapunov analysis}

The average auto-correlation time is only one measure of predictability of single oscillator trajectories. Complementary measures can be defined by Lyapunov exponents, which quantify the average rate of divergence between nearby trajectories. There are $d$ possible directions in a $d$-dimensional system and thus up to $d$ different Lyapunov exponents describing expanding and contracting subspaces \cite{oseledec1968multiplicative}. Positive Lyapunov exponents $\lambda>0$ indicate subspaces in which an initial deviation from the trajectory grows exponentially in time at rate $\lambda$. Phase space volume on a chaotic attractor, e.g. a small box in the expanding directions around a reference trajectory which co-evolves with the flow, grows on average exponentially at a rate given by the sum of the positive Lyapunov exponents. The logarithm of that phase space volume is an entropy which quantifies uncertainty in the sense of information theory. Its growth rate is the sum of the positive Lyapunov exponents, the dynamic Kolmogorov-Sinai entropy, which in coupled systems is often extensive in the system size \cite{pikovsky2016lyapunov}. If we divide the Kolmogorov-Sinai entropy by the system size $N$ we obtain the intensive quantity
\begin{equation}
    h_{KS} = \frac{1}{N}\sum_{n=1, \lambda_n>0}^{d} \lambda_n.
\end{equation}
This is the contribution of each unit to the growth rate of uncertainty in the whole coupled system and may be taken as a complementary measure to the auto-correlation time. While the latter is a rather global description of predictability, $h_{KS}$ is a local description connected to how fast a localized initial set of trajectories diverge on average.
The system of $N$ coupled Kuramoto oscillators with inertia which we are considering is $2N$-dimensional and we can determine its $d=2N$ Lyapunov exponents the usual way \cite{oseledec1968multiplicative,shimada1979numerical,benettin1980lyapunov,pikovsky2016lyapunov} by integrating the linearized dynamics along a system trajectory. Suppose $\vartheta_n(t)$ for $n=1\ldots N$ is a solution of \eqref{eq:eom} and $h_n(t)$ are small deviations from that solution. Let $v_n=\dot{h}_n$ be the velocities of the small deviations relative to the reference solution. Then in linear order the dynamics of the $v_n(t)$ is given \cite{olmi2015chimera} as 
\begin{equation}
    m\dot{v}_n = -v_n + \sum_{m=1}^N \mathrm{Re}\left[e^{-i\vartheta_n}(K_{nm}e^{i\vartheta_m}-\delta_{nm}\zeta_n(t))\right]h_m.
\end{equation}
The linear dynamics of the small deviations is forced by the full dynamics of the reference solution $\vartheta_n(t)$.
In each time step we integrate in $\Delta t$ the $2N$ independent column vectors of a matrix $\mathrm{M}(t,\Delta t)$ representing $2N$ independent small deviations $h_n(t+\Delta t)$ and $v_n(t+\Delta t)$ starting from $\mathrm{M}(t,0)=\mathbf{1}$ using the Runge-Kutta RK4 method for the phases $\vartheta_n(t+\Delta t)$ and for $\mathrm{M}(t,\Delta t)$. The matrix $\mathrm{M}(t,\Delta t)$ is the linearized $\Delta t$-forward map along the reference trajectory. That is, $h_n(t+\Delta t)$ and $v_n(t+\Delta t)$ are given by the application of $\mathrm{M}(t,\Delta t)$ to any vector of small deviations $h_n(t)$ and $v_n(t)$. Repeated application of that linear map to a co-moving, orthogonal matrix $Q(t)$ and subsequent QR decomposition of $\mathrm{M}(t, \Delta t) \mathrm{Q}(t) = \mathrm{Q}(t + \Delta t) \mathrm{R}(t)$ yields, after a sufficiently long transient, the Oseledec vectors in the columns of $\mathrm{Q}(t)$.  The local Lyapunov exponents in units of inverse time are the logarithm of the absolute values of the diagonal elements of $\mathrm{R}(t)$ divided by the time step $\Delta t$
\begin{equation}
    \lambda_n(t) = \lim_{\Delta t\to 0} \frac{1}{\Delta t}\log |R_{nn}(t)|.
\end{equation}
The time averages of the local Lyapunov exponents along the trajectory are the Lyapunov exponents
\begin{equation}
    \lambda_n = \lim_{T\to\infty}\frac{1}{T}\int_{t_0}^{t_0+T}\lambda_n(t)\,dt
\end{equation}
after a sufficiently long transient $t_0$ which is necessary for the Oseledec vectors to become independent from an arbitrary initial matrix $\mathrm{Q}(0)$.
We use the positive Lyapunov exponents to measure the Kolmogorov-Sinai entropy per oscillator $h_{KS}$ for several random network realizations in systems of $N=100$ coupled oscillators with inertia, zero mean field coupling $K=0$, unit coupling heterogeneity $k=1$ and without frequency disorder, i.e. $\sigma_\omega=0$. As shown in \bi{Lyapunov}a  $h_{KS}$ grows from a small value at zero mass and reaches a maximum around $m=8$. While the number of positive Lyapunov exponents increases for increasing $m$ (\bi{Lyapunov}b), the Lyapunov exponents move closer to zero, indicating a slowing down of the dynamics, so that the sum of the positive Lyapunov exponents decreases again for $m>8$. \\
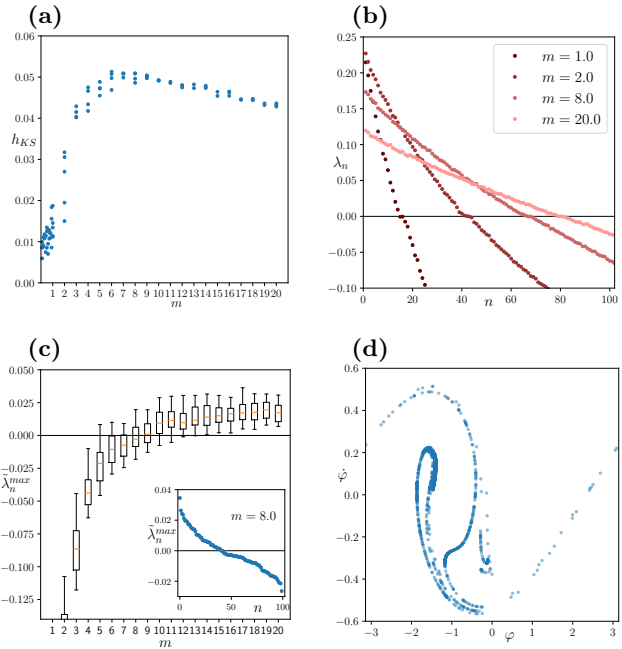
\begin{figure}[t!]
\setlength{\unitlength}{1cm}
\begin{picture}(4.1,4.4)
\put(-0.2,0){\includegraphics[width=0.215\textwidth]{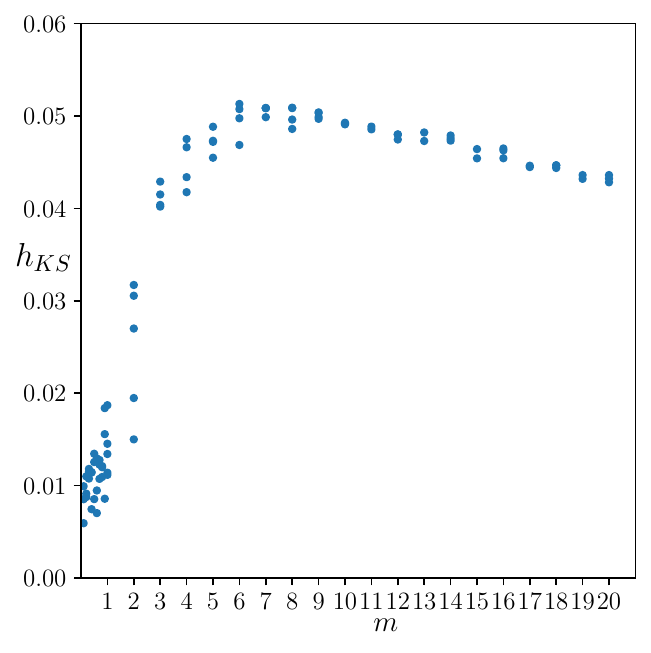}}
\put(0.1,3.9){\bf (a)}
\end{picture}
\begin{picture}(4.1,4.6)
\put(-0.2,0){\includegraphics[width=0.225\textwidth]{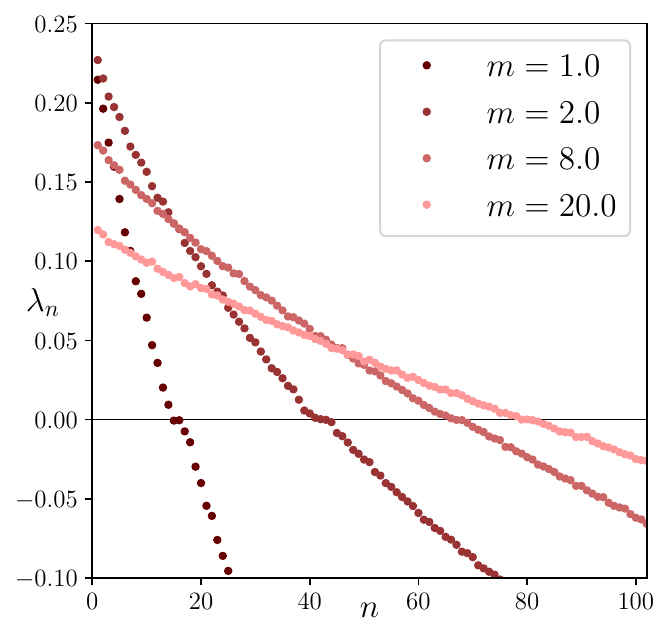}}
\put(0.2,3.9){\bf (b)}
\end{picture}
\begin{picture}(4.1,4.4)
\put(-0.35,-0.1){\includegraphics[width=0.225\textwidth]{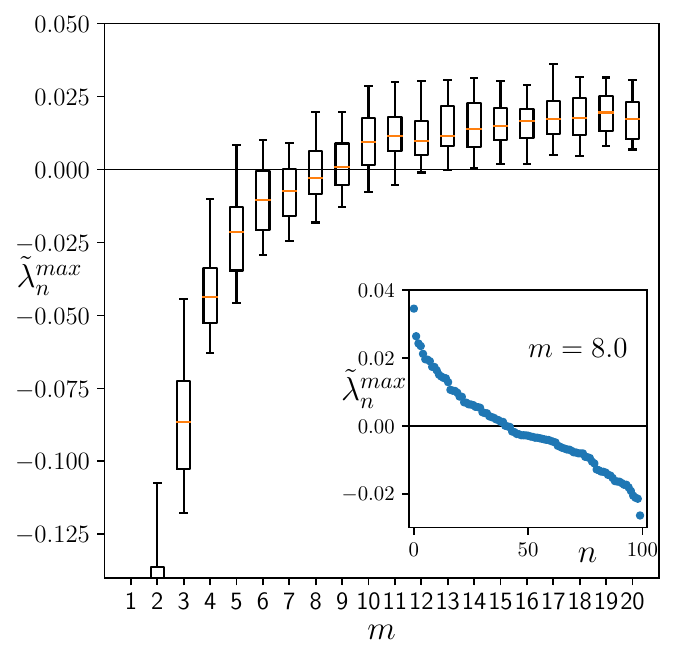}}
\put(0.1,3.9){\bf (c)}
\end{picture}
\begin{picture}(4.1,4.4)
\put(-0.1,0){\includegraphics[width=0.22\textwidth]{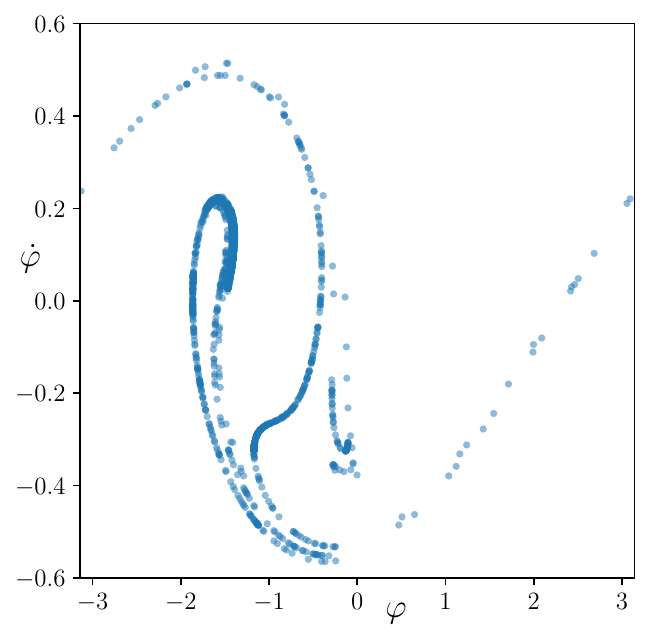}}
\put(0.2,3.9){\bf (d)}
\end{picture}
\caption{\textbf{Lyapunov analysis of the randomly coupled Kuramoto model with inertia.} Shown are results for single network realizations, zero frequency disorder $\sigma_\omega=0$, unit coupling heterogeneity $k=1$ and no mean field coupling $K=0$. (a) Kolmogorov-Sinai entropy $h_{KS}$ per oscillator calculated as the sum of positive Lyapunov exponents divided by the number of oscillators (maximum around $m=8$). For each $m$ we plot $h_{KS}$ for several network realizations to indicate the finite-size variability. (b) Lyapunov exponents for different masses, each for a single but different realization of the random network. (c) Plot of single-oscillator maximum Lyapunov exponents as function of the mass. The Lyapunov exponents are distributed. Shown are boxes over the center quartiles 25\% to 75\%, the median and bars over the range 5\% to 95\%. The inset shows all single-oscillator maximum Lyapunov exponents in one realization with mass $m=8.0$. (d) Snapshot attractor of $2000$ tracer oscillators with mass $m=8.0$ forced by a local mean field $\zeta(t)$ where the single oscillator Lyapunov exponent assumes the positive value $\tilde{\lambda}_n^{max}=0.023$.}
\label{f:Lyapunov} 
\end{figure}
Instead of measuring the $2N$ Lyapunov exponents $\lambda_n$ of the $2N$-dimensional coupled system we can also look at the single oscillators, forced by the complex local mean fields $\zeta_n(t)$. We can define two single-oscillator Lyapunov exponents $\tilde{\lambda}_n^{low}$ and $\tilde{\lambda}_n^{max}$ by integrating $N$ $2\times 2$ matrices $\mathrm{M}_n(t,\Delta t)$ where the linearized equations of motion in addition to $\dot h_n = v_n$ are
\begin{equation}
    m\dot v_n = -v_n -\mathrm{Re}\left[e^{-i\vartheta_n}\zeta_n(t)\right]h_n.
\end{equation}
While the lower Lyapunov exponent is negative due to the dissipation, the second Lyapunov exponent can be either negative or positive and is in general distributed for different oscillators in the random network. If it is positive, the forced dynamics of the corresponding single oscillator is chaotic. The broadening of the power spectra for intermediate mass may be a signature of this chaotic single-oscillator dynamics. In \bi{Lyapunov}c we show the distribution of the single-oscillator maximum Lyapunov exponents for single network realizations and zero frequency heterogeneity $\sigma_\omega=0$ as a function of mass. For $m\to 0$ and $\sigma_\omega=0$ all single-oscillator Lyapunov exponents are smaller than zero. Increasing the inertia the Lyapunov exponents grow and start to become positive for $m>4$. At $m=8$ about half of the oscillators have become chaotic. Since the maximum Lyapunov exponent of an oscillator describes the divergence of nearby trajectories from the trajectory of the reference oscillator, the transition to chaotic dynamics can be visualized through the emergence of a chaotic snapshot attractor in tracer oscillators $\varphi_k$ forced by one realization of the complex local mean field $\zeta_n(t)$ of the chaotic reference oscillator
\begin{equation}
    m\ddot{\varphi}_k = \omega_n -\dot{\varphi}_k + \mathrm{Im}\left[e^{-i\varphi_k}\zeta_n(t)\right].
\end{equation}
If $\tilde{\lambda}_n^{max}<0$ deviations from the reference oscillator vanish and the tracers synchronize to the reference oscillator $\varphi_k(t)\to \vartheta_n(t)$. On the other hand, if $\tilde{\lambda}_n^{max}>0$ the trajectories of the tracers diverge and fill a chaotic snapshot attractor (\bi{Lyapunov}d). 
\section{Conclusion}
\label{sec:conclusions}
We have applied the iterative mean-field (IMF) method to obtain the self-consistent power spectra of oscillators and network noise in a Kuramoto model with inertia and disorder in natural frequencies and connections. Our results from the IMF method show a strong agreement with the numerical outcomes obtained by solving the full $N$-dimensional dynamics whenever the system is in an asynchronous state. 

Both by network simulations and by the IMF method we observed a remarkable effect, namely, that the bandwidth of the power spectral peak is maximized and, correspondingly, the correlation time of the oscillators is minimized for an intermediate mass of the oscillators. The effect is observed when the correlation function turns from a purely decaying function into an oscillating one (with a decaying envelope). We also found that around the optimal mass which minimizes temporal correlations, the dynamic Kolmogorov-Sinai entropy in the network is maximized. We traced this back to the structure and distribution of the Lyapunov exponents, both the exponents for the full (high-dimensional autonomous deterministic) system and for the single oscillators (low-dimensional and driven by an effective noise). 

Is there any hope for an analytical approach to capture the minimization of correlations in the self-consistent fluctuations of a disordered Kuramoto model with inertia? For a similar system without inertia, Pr\"user et al. \cite{PruRos24,PruEng24} worked out a weak-coupling approximation for the self-consistent correlation function that provides explicit expressions for the correlation function (the inverse Fourier transform of the spectra considered in our study for $m=0$). Their model also includes a possible correlation or anti-correlation between the coupling coefficients $K_{\ell m}$ and $K_{m \ell}$ and the authors focused on the Volcano transition \cite{Dai93,OttStr18}. They show rather good agreement between the correlation functions as predicted by their weak-coupling theory, numerical simulations of the network, and an single-oscillator simulation scheme, equivalent to what we referred to as the IMF method. Their theory is based on the smallness of the coupling and constitutes an expansion of correlation as well as response functions (the latter are needed when the coupling coefficients are not independent) in terms of the small parameter $k/\sigma_\omega\ll 1$ (corresponding to small $J$ in the nomenclature used in \cite{PruRos24,PruEng24}). Unfortunately, even if the theory can be generalized to include inertia, the effect of minimized correlation for intermediate mass requires a strong coupling, i.e. $k/\sigma_\omega\gg 1$ and thus even a generalized perturbation theory is unlikely to capture the observed effect.

What are the implications of minimized correlation times?  Temporal correlations of intrinsic fluctuations shape the signal transmission and processing properties of a network (see e.g. the discussion in \cite{Ost14}). Thus, the mass parameter may not only be optimal with respect to a short correlation time but also most or least favorable with respect to the processing of information about time-dependent stimuli. For instance, if, for a network with time-dependent driving, there are simple relations between the response functions (or their Fourier counterpart, the susceptibility)  and the correlation function (or their Fourier-version, the power spectrum), a high cutoff frequency of the spontaneous spectra may imply that the network's susceptibility has a high cutoff frequency as well, permitting the transmission of fast stimuli. The latter feature is, for instance, important to understand how fast signal transmission in the brain can take place, specifically, the surprisingly short reaction times to brief visual stimulation. 

Relations between the correlations of spontaneous fluctuations (no stimulus) and the response functions with respect to time-dependent signals are known as fluctuation-dissipation theorems or fluctuation-response relations (FRRs) \cite{MarPug08}. Such relations have been found more recently for a number of non-equilibrium scenarios with stochastic oscillators \cite{PerGut23}, spiking neurons \cite{Lin22,PutLin24,StuLin24}, and neural populations \cite{SarArv20,DecLyn23,NanCan23}.  For the disordered Kuramoto model \emph{without} inertia, Pr\"user and Engel obtained a simple FRR in which the response function is simply proportional to the correlation function for the case of weak antisymmetric coupling ($K_{\ell m}=-K_{m \ell}$ in terms of our parameters). It is an interesting challenge for future research to tackle the case with inertia and completely random coupling coefficients, including in our model studied here time-dependent perturbations of the oscillators and to derive relations between response functions and spontaneous correlations for this system.  The explicit calculation of the self-consistent response functions in parallel with the self-consistent correlation functions, as put forward by Pr\"user et al. in the system without inertia may also enable us to calculate the signal-to-noise ratio and ultimately estimates on the mutual information rate of the network.

\section*{Acknowledgments}
This work was funded by the German Research Foundation (DFG) - 506198590. The code used for our computations is available at \cite{Kati2025Power}.

\appendix

\section{Time rescaling}
\label{app:rescaling}
Suppose we use $\tau$ as the new unit of time $\bar{t}$, i.e. $\bar{t} = t/\tau$. Then the first and the second derivatives with respect to $t$ are replaced by the corresponding derivatives with respect to $\bar{t}$ and a prefactor of $\tau^{-1}$ and $\tau^{-2}$ respectively and the rotator model with $K=0$ becomes
\begin{equation}
    m\frac{1}{\tau^2}\frac{d^2}{d\bar{t}^2} \theta_\ell = \gamma\left(\omega_\ell-\frac{1}{\tau}\frac{d}{d\bar{t}}\theta_\ell\right) +
    \frac{k}{\sqrt{N}}\sum_{m=1}^N \mathcal{G}_{\ell m}\sin(\theta_m-\theta_\ell)
\end{equation}
After dividing by $k$ and setting $\tau=\gamma/k$, the dynamics is given by
\begin{equation}
    \bar{m}\frac{d^2}{d\bar{t}^2}\theta_\ell = \left(\bar{\omega}_\ell-\frac{d}{d\bar{t}}\theta_\ell\right)+ 
    \frac{1}{\sqrt{N}}\sum_{m=1}^N \mathcal{G}_{\ell m}\sin(\theta_m-\theta_\ell)
\end{equation}
where the frequencies $\bar{\omega}_\ell=\tau\omega_\ell$ are now in units of $1/\tau$ and the effective mass is $\bar{m}=mk/\gamma^2$. All temporal observables can be obtained by a rescaling of a measurement with a given $\bar{m}$ and frequency heterogenity $\bar{\sigma}_\omega=\tau\sigma_\omega$. If, for instance with $\sigma_\omega=0$, the autocorrelation time $\bar{\tau}_C$ is known as a function of $\bar{m}$, then the autocorrelation time $\tau_c(m,\gamma,k)$, stating its explicit dependence on $m$, $k$ and $\gamma$, is related to that for $k=1$ and $\gamma=1$ by 
\be
\tau_c(m,\gamma,k) = \frac{k}{\gamma}\tau_c(mk/\gamma^2,1,1)=\frac{k}{\gamma}\bar{\tau}_C(\bar{m}).
\label{eq:tauc_rescaled_gamma}
\ee
The frequency band is rescaled as (again for the case $\sigma_\omega=0$)
\be
    \Delta\omega(m,\gamma,k) = \frac{\gamma}{k}\Delta{\omega}(mk/\gamma^2,1,1)=\frac{\gamma}{k}\Delta{\bar{\omega}}(\bar{m}).
    \label{eq:del_om_rescaled_gamma}
\ee
Analog relations hold true for characteristic frequencies, Lyapunov exponents etc. 
Note that a non-vanishing frequency disorder, $\sigma_\omega>0$, requires a rescaling of $\sigma_\omega>0$ too when $\gamma$ or $k$ are changed:
\be
    \sigma_\omega(m,\gamma,k) = \frac{\gamma}{k}\sigma_\omega(mk/\gamma^2,1,1).
    \label{eq:sigma_om_rescaled_gamma}
\ee

\bibliographystyle{apsrev4-2}
\bibliography{bib1,bib2}

\end{document}